\documentclass[12pt]{article}
\usepackage{bm,dcolumn,amsmath,graphicx,amsfonts,amssymb}

\newcommand{\degree}{^{\circ}}

\usepackage{times}

\newenvironment{sciabstract}{%
\begin{quote} \bf}
{\end{quote}}

\title{Optical clock comparison test of Lorentz symmetry}

\author
{Christian Sanner$^{1\ast}$, Nils Huntemann$^{1}$, Richard Lange$^{1}$, Christian Tamm$^{1}$, \\ Ekkehard Peik$^{1}$, Marianna S. Safronova$^{2,3}$ \& Sergey G. Porsev$^{2,4}$\\
\\
\normalsize{$^{1}$Physikalisch-Technische Bundesanstalt, Bundesallee 100, 38116 Braunschweig, Germany}\\
\normalsize{$^{2}$Department of Physics and Astronomy, University of Delaware, Newark, DE 19716, USA}\\
\normalsize{$^{3}$Joint Quantum Institute, National Institute of Standards and Technology}\\
\normalsize{and the University of Maryland, College Park, MD 20742, USA}\\
\normalsize{$^{4}$Petersburg Nuclear Physics Institute of NRC "Kurchatov Institute", Gatchina 188300, Russia}\\
\\
\normalsize{$^\ast$Present address: JILA, 440 UCB, Boulder, CO 80309, USA}
}

\date{}

\begin{document}

\baselineskip24pt

\maketitle

\begin{sciabstract}

Questioning the presumably most basic assumptions about the structure of space and time has revolutionized our understanding of Nature.
State-of-the-art atomic clocks
make it possible to precisely test fundamental symmetry properties of spacetime, and
search for physics beyond the standard model
at low energy scales of just a few electron volts.
Here, we experimentally demonstrate for the first time agreement of two single-ion clocks at the
$10^{-18}$ level and directly confirm
the validity of their uncertainty budgets
over a half-year long comparison period.
The two clock ions are confined in separate ion traps with
quantization axes aligned along nonparallel directions.
Hypothetical Lorentz symmetry violations
would lead to sidereal modulations of the frequency offset.
From the absence of such modulations at the $10^{-19}$ level we deduce stringent limits on
Lorentz symmetry violation parameters
for electrons in the range of $10^{-21}$, improving previous limits by two orders of magnitude.
\end{sciabstract}
The principle of relativity requires that all descriptions of Nature are covariant under Lorentz transformations, i.e., the laws of physics stay the same when transforming from one inertial reference frame to another. Consequently, the outcome of any experiment must be independent of the velocity and orientation of the inertial frame in which it is performed.
The Michelson-Morley experiment \cite{Michelson1887} with a rotatable optical interferometer was an early test of this observational symmetry, disproving the existence of a preferred reference frame orientation for electromagnetic waves. Later, Kennedy and Thorndike \cite{Kennedy1932} modified the Michelson-Morley setup to also establish velocity invariance. Strong interest persists in improved tests of Lorentz symmetry in different branches of physics, motivated by theoretical suggestions that local Lorentz covariance may not be an exact symmetry at
all energies up to the Planck energy and is violated in various models of quantum gravity \cite{Mattingly2005}.
First spectroscopic tests of Lorentz symmetry have been performed in nuclear magnetic resonance, when a dependence of the magnetic splitting of energy levels of the $^7$Li nucleus on its orientation relative to the center of our galaxy was excluded at a level of about 1 ppm \cite{Hughes1960,Drever1961}. The apparatus was fixed in the laboratory, rotating with Earth once per sidereal day. Related experiments have since been done with different nuclei, providing sensitivity to Lorentz violations (LV) for protons and neutrons.
LV limits for photons have been improved continually through numerous tests with high-finesse optical cavities (see Ref. \cite{Safronova2018} for a review). Such cavity measurements also lead to LV constraints for electrons \cite{Mueller2003,Mueller2005}, which have been further advanced by astrophysical observations \cite{Altschul2006} and more recently by direct measurements with bound electrons in Dy atoms \cite{Hohensee2013} and Ca$^+$ ions \cite{Pruttivarasin2015}.

In parallel, the development of optical clocks has seen tremendous progress over recent years.
Laser spectroscopy with $10^{-18}$ fractional frequency accuracy \cite{Nicholson2015,Huntemann2016} and coherence times reaching tens of seconds is now within experimental reach \cite{Campbell2017}.
We demonstrate in our experiment
an unprecedented level of performance of single-ion clocks by showing persistent
agreement of two $^{171}$Yb$^+$ systems at the low $10^{-18}$ level
over the course of half a year. Long term comparisons between
such excellent quantum timekeepers
open new avenues for low-energy tests of fundamental physics \cite{Safronova2018}.
Here, we exploit
the pronounced LV susceptibility of the non-spherical $^{2}$F$_{7/2}$ clock state in Yb$^+$ \cite{Dzuba2016} and
implement a high-precision spacetime anisotropy test
with two ion clocks operating on the same optical transition but being oriented along different quantization axes.
To systematically quantify our frequency measurements in terms of LV constraints we rely on a theoretical framework called Standard Model Extension (SME)  which provides a universal platform to compare all kinds of LV measurements \cite{Colladay1998}.
The SME covers all standard model particles and can therefore identify LV in all branches
of the standard model.

Considering
the electron sector, a hypothetical LV is quantified in the SME by adding a symmetry-breaking $c_{\mu \nu}$ tensor to the kinetic term in the standard model Lagrangian \cite{Kostelecky1999} . In the context of our clock comparison experiment,
this leads
to a non-common-mode energy shift of bound electronic states described by the Hamiltonian~\cite{Kostelecky1999,Hohensee2013}
\begin{equation}
\delta H =
-C_{0}^{(2)} \, \frac{\mathbf{p}^{2} - 3p_{z}^{2}}{6 m} \equiv
-C_{0}^{(2)} \, T_{0}^{(2)} / \, 6m,
\label{eq1}
\end{equation}
where $\mathbf{p}$ is the momentum of a bound electron, $p_{z}$ its projection along the quantization axis, and $m$ the electron mass.
The parameter
$C_{0}^{(2)}=c_{xx} + c_{yy} - 2 c_{zz}$ contains spatial elements of the frame-dependent $c_{\mu \nu}$ tensor and
$T_{0}^{(2)} $ is the corresponding component of the quadrupole moment operator
of the electronic momentum distribution.
However, instead of analyzing our experiment in the electron sector and deriving anisotropy limits for the electronic dispersion, one can also 
base the analysis on anisotropies in the photon-mediated Coulomb field.
Accordingly, the electron and the photon parts cannot be separated and
by redefining
\begin{equation}
c'_{\mu \nu}=c_{\mu \nu} +\frac{1}{2} k_{\mu \nu},
\label{ElPlusPh}
\end{equation}
where the two terms refer to the electron and photon LV coefficients \cite{Pruttivarasin2015,Dzuba2016}, respectively, the present constraints describe differential electron-photon anisotropies and can be interpreted to limit either sector.
To simplify notation we omit primes from now on
and always assume sector-combined tensor components.
In atoms like $^{171}$Yb with nuclear spin $I=1/2$, LV in the nuclear sector cannot produce quadrupole energy shifts.

\begin{figure}[p]
\centering
\includegraphics[scale=0.6]{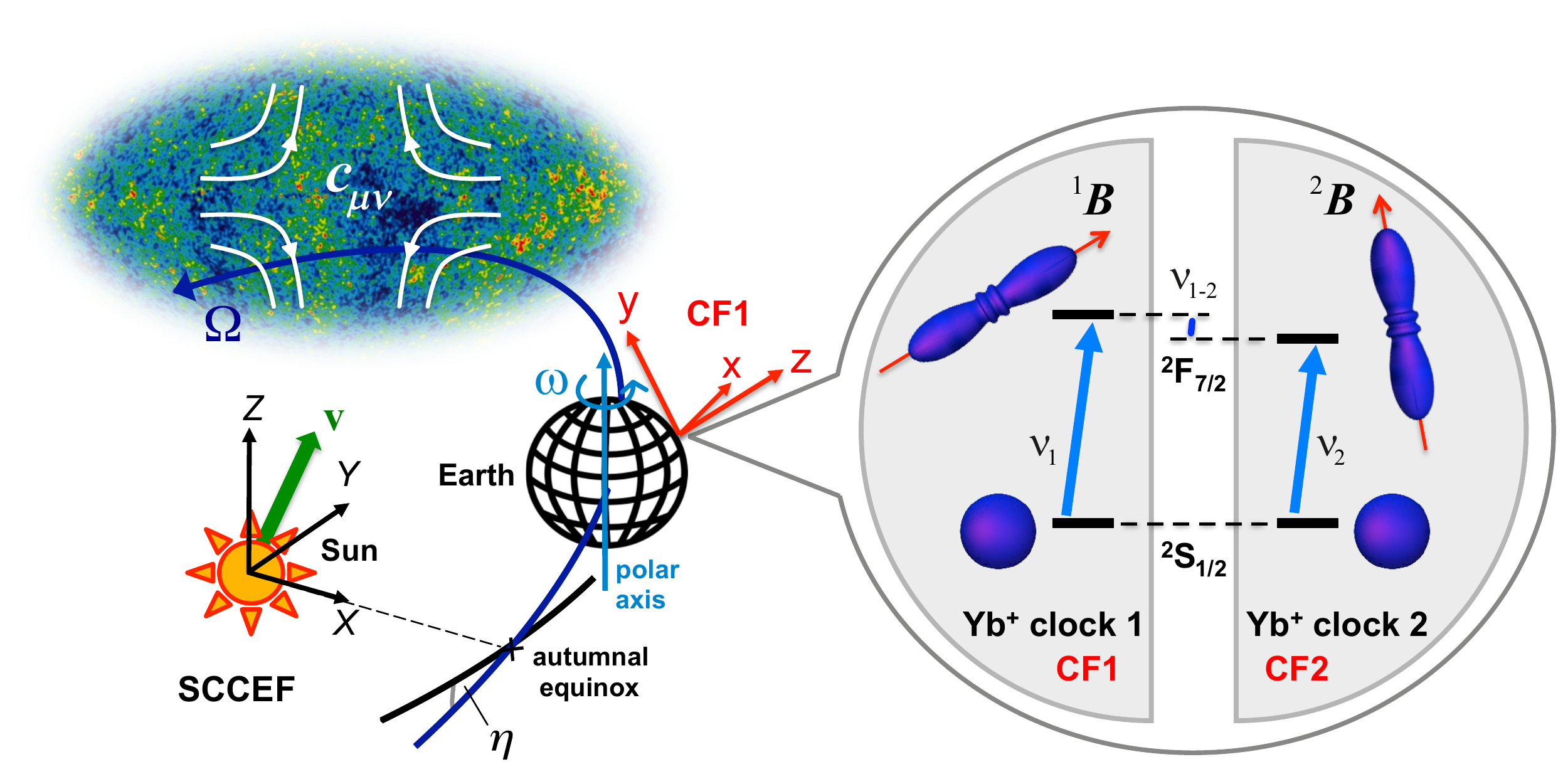}
\caption{Testing Lorentz symmetry with two earthbound optical Yb$^+$ clocks. Since the sun-centered celestial-equatorial frame (SCCEF) translates with approximately constant velocity \textbf{v} relative to the cosmic microwave background rest frame \cite{Bluhm2003}, we assume that the LV parameter tensor $c_{\mu \nu}$ is constant in the SCCEF. The clock frames CF1 and CF2 (only CF1 is shown) have their $z$-axes aligned along the respective atomic quantization axes defined by the magnetic fields $\bm{^1B}$ and $\bm{^2B}$. Earth rotates with angular frequency $\omega$ about its polar axis and orbits the Sun with $\Omega$, so that the SCCEF-referenced orientations and kinetic boosts of the two clock frames acquire corresponding sidereal time dependences. A Lorentz violation in the electron-photon sector leads to frame-dependent atomic transition frequencies. Specifically, the upper $^{2}$F$_{7/2}$ $(F=3, m_F=0)$ clock state with its anisotropic electron momentum distribution \cite{FootNote1} experiences a differential energy shift between CF1 and CF2 that varies as Earth rotates. This will cause a corresponding modulation of the frequency difference $\nu_{1-2}$ between the two clocks that can be measured in the laboratory.}
\label{EarthSunStars}
\end{figure}

Our experiment aims at measuring the energy shift  $\langle \delta H \rangle$
by comparing the frequencies $\nu_1$ and $\nu_2$ of two $^{171}$Yb$^{+}$ single-ion clocks operating on the $^{2}$S$_{1/2}$ - $^{2}$F$_{7/2}$
electric octupole (E3) transition as illustrated in Figure~\ref{EarthSunStars}. The two ions are
stored in separate Paul traps \cite{Ludlow2015} within the same laboratory. Differential sensitivity to LV induced shifts arises from different magnetic field orientations $\bm{^1B}$ and $\bm{^2B}$ in the two ion traps. The magnetic field defines the quantization axis for each ion and lifts the $m_F$ degeneracy within the excited state $^{2}$F$_{7/2}$ $(F=3)$ hyperfine manifold. Coupling of the nuclear spin $I=1/2$ and the electron angular momentum $J=7/2$ leads to $m_F = 0$ states
that are first-order insensitive to magnetic field
noise. The latter presented a major challenge for previous \cite{Safronova2018} and recently proposed LV searches \cite{Dzuba2016}, where field fluctuations had to be addressed by engineering decoherence-free subspaces \cite{Roos2006}.
Unlike the spherical $^{2}$S$_{1/2}$ $(F=0, m_F=0)$ ground state, the excited $^{2}$F$_{7/2}$ $(F=3, m_F=0)$
clock state has an anisotropic momentum distribution \cite{Dzuba2016,SupplMat} with an exceptionally large expectation value
$\langle T_{0}^{(2)} \rangle / \, 6m \equiv h q  = h \times 2.6 \times 10^{16}$~Hz, where $h$ is the Planck constant,
and will experience a LV shift proportional to each ion's local $C_{0}^{(2)}$ value which we label $^{1}C_{0}^{(2)}$ and $^{2}C_{0}^{(2)}$. These values will differ for different quantization orientations and will change as Earth rotates. The $c_{\mu \nu}$ tensor is uniquely specified \cite{Ding2016,KostVargas2018} in the Sun-centered celestial-equatorial frame (SCCEF) and Lorentz transformations $\bm{^{1}\Lambda}$ and $\bm{^{2}\Lambda}$ provide mapping between the SCCEF (upper case indices) and the respective clock frames CF1 and CF2 (lower case indices), i.e.,
\begin{equation}
^{1,2}c_{mn} = {^{1,2}\Lambda}_m^{\;M} \;\; {^{1,2}\Lambda}_n^{\;N}
\;\, c_{MN}.
\label{LTs}
\end{equation}
By continuously measuring
the frequency difference $\nu_{1-2} = \nu_1 - \nu_2$ between the two clocks and relating it to the instantaneous orientations and kinetic boosts of CF1 and CF2 with respect to the SCCEF, we can derive upper bounds for the $c_{MN}$ components.

While several atomic clocks \cite{Nicholson2015,Huntemann2016,Ludlow2015} have been characterized with systematic fractional uncertainty budgets below $10^{-17}$, it remains a formidable challenge to prove
this level of accuracy in actual clock comparison experiments. This particularly applies to ion clocks, where quantum projection noise associated with the spectroscopic interrogation of a single atom limits the attainable frequency stability \cite{Ludlow2015}.
Here, we present a half-year long
frequency comparison between two optical $^{171}$Yb$^+$ single-ion clocks,
showing an agreement of the measured 642.121~THz electric octupole transition frequencies
to better than 2~mHz. This corresponds to
a relative frequency offset of $2.8 \times 10^{-18}$, measured with an unparalleled $2.1 \times 10^{-18}$ statistical uncertainty.

Relying on the same E3 reference transition, our two $^{171}$Yb$^+$ clocks
obtain laser light to drive the clock transition via individual acousto-optic frequency shifters
from the same silicon-cavity-prestabilized laser source \cite{Matei2017}, but
have otherwise substantial differences in the experimental implementation.
Table~\ref{Shifts} summarizes the leading systematic frequency shifts with corresponding uncertainties
for both clocks. In reference \cite{Huntemann2016}
we provide
a complete discussion of all shift effects contributing to the uncertainty budget. Additional details regarding the diagnosis and correction of clock shifts are found in \cite{SupplMat}.
For clock 1, the ion is stored in a spherically symmetric Paul trap with a central ring electrode
providing confinement with radial secular frequencies of $^1\omega_r = 590$~kHz, while clock 2 employs an end-cap trap design with $^2\omega_r = 1060$~kHz. Since both traps store a single ion for several months, clock operation is not
affected by loading-related perturbations.

\begin{table}[t]
\centering
{\begin{tabular}{lcccc}
Shift effect & $\Delta\nu_1/\nu_0\: (10^{-18})$ & $\Delta\nu_2/\nu_0 \: (10^{-18})$ & $\Delta\nu_{1-2} /\nu_0\: (10^{-18})$  \\ \hline
Second-order Doppler    & -2.3 (1.5)    &   -4.0 (2.2)  & 1.7 (2.7)   \\
Blackbody radiation     & -70.5 (1.8)   & -69.9 (1.7)   & -0.6 (1.7)  \\
Probe light             &  0 (0.8)      & 0 (0.5)       &   0 (0.9) \\
Second-order Zeeman     & -10.4 (0.2)   &  -11.3 (0.4)  &  0.9 (0.4)   \\
Quadratic dc Stark      & -0.8 (0.6)    & -1.3 (0.8)    & 0.5 (1.0)   \\
Quadrupole              & -5.7 (0.5)    & -3.9 (0.5)    & -1.8 (0.7)   \\
Background gas          & 0 (0.5)       & 0 (0.5)       & 0 (0.7) \\
Servo                   & 0 (0.2)       & 0 (0.1)       & 0 (0.1)\\  
Gravitation             & -0.5 (0.1) & 0 & -0.5 (0.1)  \\ \hline
Total                   & -90.2 (2.7) & -90.4 (3.0) & 0.2 (3.6)
\end{tabular}}
\caption{Systematic frequency shifts $\Delta \nu_{1,2}$ for clocks 1 and 2 and corresponding uncertainties given as fractions of the unperturbed E3 transition frequency $\nu_0$. Differential shift uncertainties for $\nu_{1-2} = \nu_1 - \nu_2$ can be smaller than expected from the individual margins if the specific sources of error, e.g., atomic polarizability uncertainties in the context of thermal radiation shifts, are common for both clocks. The gravitational redshift is specified relative to clock 2.}
\label{Shifts}
\end{table}

Given its exceedingly small oscillator strength, the excitation on the E3 transition requires very large optical intensities such that off-resonant coupling to other levels causes a non-negligible light shift of the reference transition frequency during atom-light interactions. For a 60 ms $\pi$-pulse excitation, this shift amounts to 120~Hz, nearly $10^5$ times larger than the absolute uncertainty of the clock. To address this detrimental effect, various powerful composite pulse Ramsey interrogation approaches have been explored over recent years \cite{Huntemann2012,Sanner2018,Zanon2018}. In the comparison reported here, we use interleaved two-loop Rabi / Ramsey spectroscopy schemes \cite{Huntemann2016}, where Rabi interrogations are performed to determine the position of the light-shifted transition. Clock 1 alternates its Rabi measurements with accordingly tuned hyper-Ramsey interrogations \cite{Yudin2010}, while clock 2
uses a
Ramsey protocol with frequency-shift keying between drive-pulse and dark-time intervals \cite{Taichenachev2010}. The clocks use dark Ramsey times of 400 and 360~ms, respectively, and have effective overhead times of 470 and 230~ms per interrogation that include the Rabi sequences. This implies that the clocks operate simultaneously in an asynchronous fashion.

\begin{figure}[p]
\centering
\includegraphics[scale=0.7]{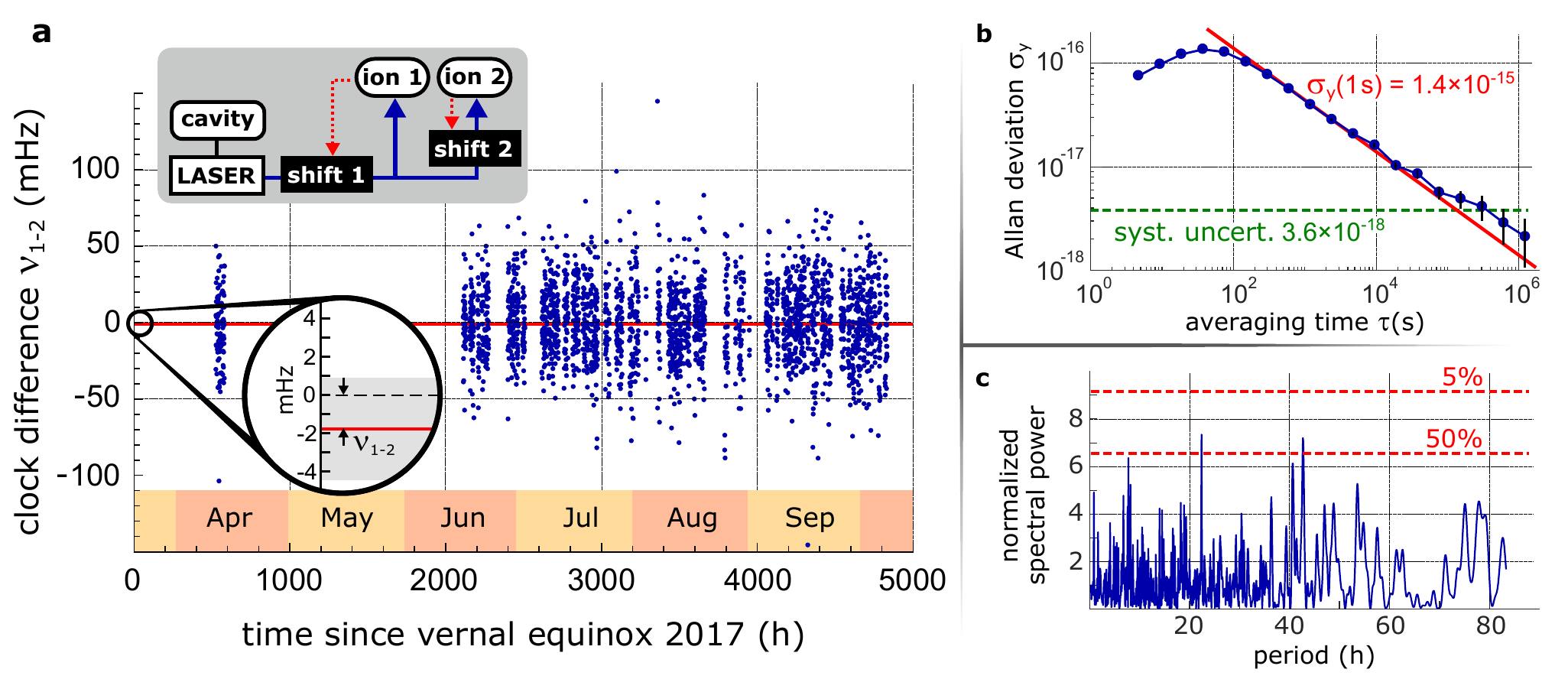}
\caption{Half-year long frequency comparison between two ytterbium single-ion clocks operating on the 642~THz electric octupole transition. \textbf{(a)} Data collected during the measurement campaign is binned into half-hour long intervals and the mean clock frequency difference for each filled bin is shown as a blue data point. The inset illustrates the chained configuration of the two clocks. Ion 1 controls a common-mode frequency shift 1, and shift 2, which is controlled via feedback from ion 2, directly reflects the frequency difference between the clocks. The plotted $\nu_{1-2}$ values and the average frequency deviation (red line) have been corrected for the systematic offsets listed in Table~\ref{Shifts}.  The magnified region shows the clock difference of $1.8(2.7)$~mHz corresponding to $2.8(4.2)\times 10^{-18}$ in fractional frequncy units. \textbf{(b)} Allan deviation derived from unbinned concatenated frequency data. After an effective averaging time of $3 \times 10^5 \, \mathrm{s}$ the clocks have reached a statistical fractional frequency deviation that matches their combined systematic uncertainty of $3.6 \times 10^{-18}$. The error bars indicate $1\sigma$ confidence intervals for the measured frequency noise. \textbf{(c)} A Lomb normalized periodogram enables spectral analysis for unevenly sampled data. The dashed lines specify significance levels, i.e., the probabilities to find one or more peaks above these lines for a corresponding uncorrelated data set are 5\% and 50\%, respectively. No significant periodic signal is observed and the clock difference remains unaffected by periodically changing environmental conditions.
}
\label{Plots}
\end{figure}

Figure~\ref{Plots} shows the data
acquired over a six month measurement period. Beside the exclusion of data points obtained during faulty
operation conditions, no data post-selection is performed. The frequency noise is investigated via Allan deviation and Lomb periodogram analyses. An averaging behavior which mainly
follows
$\sigma (\tau) = 1.4 \times 10^{-15} \; / \sqrt{\tau (\mathrm{s})}$ indicates white frequency noise close to the expected quantum projection noise limit of $1.1 \times 10^{-15} \; / \sqrt{\tau (\mathrm{s})}$ that we estimated from a simulation of the clock comparison based on the observed Ramsey fringe contrast and
assuming an undisturbed local laser oscillator. Deviations from the $1/\sqrt{\tau}$ scaling for the last few data points on the Allan deviation plot are mostly within standard deviation bounds and might just reflect the limited sampling statistics for very long averaging times. However, instead of extrapolating to our total effective averaging time of $3.9 \times 10^6$~s, we conservatively infer a statistical uncertainty for our frequency comparison from the last Allan deviation data point of $2.1 \times 10^{-18}$. This value is added
in quadrature to the systematic uncertainty of $3.6 \times 10^{-18}$ and results in a total uncertainty of $4.2 \times 10^{-18}$ for a measured mean fractional frequency difference $\nu_{1-2} / \nu_{0} = 2.8 \times 10^{-18}$ remaining after all systematic shift corrections.

In order to derive LV constraints from the acquired frequency data, we first construct a proper fit function $\nu_{1-2} (t) = -q \, [{^{1}C}_{0}^{(2)}(t) - \, {^{2}C}_{0}^{(2)}(t)]$ via equations \ref{eq1} and \ref{LTs}. The Lorentz transformation matrix $\bm{^{1}\Lambda}$ ($\bm{^{2}\Lambda}$) depends on the rotation $\bm{^1R}$ ($\bm{^2R}$) and boost $\bm{^1\beta}$ ($\bm{^2\beta}$) of CF1 (CF2) with respect to the SCCEF.
Since $\bm{^{1,2}R}$ changes with the angular frequency $\omega$ of a sidereal day and $\bm{^{1,2}\beta}$ evolves with $\Omega$ over a sidereal year, the fit function will contain terms
oscillating at $\Omega$ and $\omega$, and at mixed frequencies $2 \omega$, $\Omega \pm \omega$, and $\Omega \pm 2 \omega$. Here, we neglect boost variations due to $\omega$ because they are by a factor of $10^{-2}$ smaller than the boost due to Earth's orbital velocity, and we only keep time-dependent terms up to first order in $\beta$. Instead of sorting according to oscillation frequency we
decompose the fit function into contributions from different $c_{MN}$ components and arrive at a linear model
\begin{equation}
\nu_{1-2} (t) = c_{TX}f_1(t) + c_{TY}f_2(t)+  c_{TZ}f_3(t) + c_{XZ}f_4(t) + c_{YZ}f_5(t)+  c_{X-Y}f_6(t)
+c_{XY}f_7(t),
\label{fitfct}
\end{equation}
where $c_{X-Y}=c_{XX}-c_{YY}$.

Our recorded data sets contain the measured frequency offset between clock 1 and 2 together with UTC time tags. Frequency values are updated after four Ramsey interrogations of clock 2, i.e., every 2.36 seconds, and in total the analysis is based on 1.7 million data points
corresponding to $3.9 \times 10^6$~s of pure data time.
First, we remove the global $\nu_{1-2}$ offset and group the data into segments of about 700 points, then we fit expression \ref{fitfct} to the list of segment mean values. The reduced chi-squared $\chi_{\mathrm{red}}^2$ for this fit is 1.05, indicating
consistent assessment of the measurement uncertainty.
In Table~\ref{tabc1} we report the obtained estimates and standard errors of the individual fit parameters. Since the fit's covariance matrix contains nonzero off-diagonal entries,
these estimates are correlated and one needs to construct uncorrelated linear combinations of
$c_{MN}$ components as
provided in the lower part of the Table. Previous experiments \cite{Hohensee2013,Pruttivarasin2015} have reported their specific linear combinations and it is necessary to define a common basis to compare results from different experiments. For simplicity we choose for this purpose the set of uncombined $c_{MN}$ components and recalculate previous bounds to this basis.
\begin{table}[t]
\centering
\resizebox{1.0\textwidth}{!}
{\begin{tabular}{crrrr}
\hline \\ [-0.6pc]
\multicolumn{1}{c}{Correlated LV parameters} &
\multicolumn{1}{c}{New limits} &
\multicolumn{1}{c}{Ca$^+$ limits \cite{Pruttivarasin2015}} &
\multicolumn{1}{c}{Dy limits \cite{Hohensee2013}} &
\multicolumn{1}{c}{Astrophysical limits \cite{Altschul2006}} \\
\hline \\ [-0.6pc]
$c_{X-Y}$&
$0.9\pm1.6\times 10^{-20}$&
$-0.2\pm2.3\times 10^{-18}$&
$2.9\pm5.7\times 10^{-17}$&
$3.3\pm6.2\times 10^{-15}$
\\[0.2pc]
$c_{XY}$&
$-6.9\pm8.0\times 10^{-21}$&
$-0.8\pm1.2\times 10^{-18}$&
$0.7\pm3.6\times 10^{-17}$&
$0.0\pm3.0\times 10^{-15}$
\\[0.2pc]
$c_{XZ}$&
$1.3\pm1.3\times 10^{-20}$&
$-3.4\pm7.9\times 10^{-19}$&
$0.9\pm1.1\times 10^{-16}$&
$0.0\pm3.0\times 10^{-15}$
\\[0.2pc]
$c_{YZ}$&
$1.7\pm1.3\times 10^{-20}$&
$-1.7\pm7.1\times 10^{-19}$&
$3.1\pm6.6\times 10^{-17}$&
$-0.4\pm2.2\times 10^{-15}$
\\[0.5pc]
$c_{TX}$&
$-4.6\pm8.4\times 10^{-17}$&
&
$5.7\pm8.3\times 10^{-15}$&
$-1.5\pm5.5\times 10^{-15}$\\[0.2pc]
$c_{TY}$&
$4.8\pm8.5\times 10^{-17}$&
&
$-8.3\pm7.5\times 10^{-13}$&
$0.5\pm1.0\times 10^{-15}$\\[0.2pc]
$c_{TZ}$&
$-2.4\pm1.6\times 10^{-16}$&
&
$1.9\pm1.7\times 10^{-12}$&
$-1.0\pm3.0\times 10^{-17}$\\[0.5pc]
\hline \\ [-1.1pc]
\end{tabular}}
\centering
\resizebox{\textwidth}{!}
{\begin{tabular}{rr}
\hline \\ [-0.6pc]
\multicolumn{1}{c}{Uncorrelated linear combinations of parameters} &
\multicolumn{1}{c}{New limits} \\
\hline \\ [-0.6pc]
$0.84c_{X-Y}-0.01c_{XY}+0.50c_{XZ}+0.19c_{YZ}+9\times 10^{-5}c_{TX}-8\times 10^{-5}c_{TY}+4\times 10^{-5}c_{TZ}$&
$-0.0\pm1.1\times 10^{-20}$\\[0.2pc]
$-0.09c_{X-Y}+0.10c_{XY}-0.21c_{XZ}+0.97c_{YZ}-2\times 10^{-5}c_{TX}+4\times 10^{-5}c_{TY}+4\times 10^{-5}c_{TZ}$&
$3.8\pm9.7\times 10^{-21}$\\[0.2pc]
$0.53c_{X-Y}+0.03c_{XY}-0.84c_{XZ}-0.14c_{YZ}+3\times 10^{-5}c_{TX}-7\times 10^{-5}c_{TY}-5\times 10^{-5}c_{TZ}$&
$-2.2\pm9.3\times 10^{-21}$\\[0.2pc]
$-0.00c_{X-Y}+0.99c_{XY}+0.05c_{XZ}-0.09c_{YZ}+6\times 10^{-5}c_{TX}+5\times 10^{-5}c_{TY}-1\times 10^{-6}c_{TZ}$&
$-7.9\pm5.1\times10^{-21}$\\[0.5pc]
$0.10c_{TX}-0.07c_{TY}+0.99c_{TZ}$&
$-2.5\pm1.6\times 10^{-16}$\\[0.2pc]
$-0.49c_{TX}+0.86c_{TY}+0.11c_{TZ}$&
$3.7\pm8.5\times 10^{-17}$\\[0.2pc]
$-0.86c_{TX}-0.50c_{TY}+0.05c_{TZ}$&
$0.4\pm8.3\times 10^{-17}$\\[0.5pc]
\hline \\ [-0.6pc]
\end{tabular}}
\caption{New and existing limits on differential electron-photon LV parameters. To directly compare our results with previous limits, the independent combinations of $c_{MN}$ components given in Refs.~\cite{Pruttivarasin2015,Hohensee2013} are linearly mapped to individual component values. One standard deviation uncertainties are specified. Diagonalizing the fit's covariance matrix leads to seven uncorrelated combinations of fit parameters. The normalized combinations dissociate into mixed combinations and time-space parameter combinations because boost components of $\bm{^{1,2}\Lambda}$ that introduce the $c_{TX,TY,TZ}$ dependence are suppressed by a factor $\beta = 1 \times 10^{-4}$ compared to rotation components.}
\label{tabc1}
\end{table}

Overall, we find no compelling evidence that any of the $c_{MN}$ components are
different from zero.
As statistically expected, some of the estimates do not
reach the null result within their respective $1\sigma$ uncertainties, but all estimates are found within $2\sigma$ bounds around zero.
We improved the best existing electron limits
for the space-space $c_{MN}$ components \cite{Pruttivarasin2015} by about two orders of magnitude.
The new $c_{TX}$, $c_{TY}$, $c_{TZ}$ constraints
are compared with laboratory limits from dysprosium measurements~\cite{Hohensee2013} and with limits based on observations of high-energy astrophysical sources \cite{Altschul2006}.
For $c_{TX}$ and $c_{TY}$ we reduced the
estimate uncertainties by about a factor of 60 and 10, respectively.

The demonstrated performance of the Yb$^+$ E3 system shows that clock
accuracies beyond the $10^{-18}$ level are within experimental reach. Future precision measurements with single quantum emitters
will tremendously benefit from this potential if the measurement
period required to reach an adequate statistical uncertainty
can be further reduced.
The $^{171}$Yb$^+$ E3 transition with its nanohertz natural linewidth is an ideal candidate in this context. Ion traps with low motional heating rates \cite{Lakhmanskiy2018}
combined with next generation ultrastable lasers will make it possible to approach optical Ramsey interrogation times lasting tens of seconds. This will vastly increase the clock's frequency stability, and together with its particularly high sensitivity to changes of the fine structure constant open up the path to precise and sensitive frequency ratio measurements as discussed in searches for ultra-light scalar dark matter \cite{Safronova2018}. Even without
multi-ion entanglement resources \cite{Roos2006} it will be possible to further improve Lorentz violation bounds by following an approach \cite{Shaniv2018} that exploits differential LV sensitivities between
magnetic sublevels of a given atomic state.

\section*{Methods}

\paragraph*{Relating SCCEF and clock frames}

The elements of the $c_{\mu \nu}$ tensor are frame dependent. To allow comparisons between different experiments, their
values are generally specified in the sun-centred, celestial-equatorial frame (SCCEF), i.e., the Sun's rest frame as shown in Figure~\ref{EarthSunStars}. The origin of the SCCEF is the Sun, $Z$ is aligned with Earth's rotation axis, $X$ points in the direction of the autumnal equinox, and the
$X-Y$ plane is parallel to Earth's equatorial plane. Earth's trajectory intersects the X-Y plane under an angle $\eta = 23.4^{\circ}$ at the equinoxes.

Two perspectives
can be taken in order to
obtain SCCEF-referenced $c_{\mu \nu}$ values. One can either analyze rotating clocks within the stationary $c_{\mu \nu}$ environment of the SCCEF, or one can analyze stationary clocks within coordinate systems
with time-dependent $c_{\mu \nu}$ elements. We follow the second approach which requires knowledge of how $c_{\mu \nu}$ transforms between frames. Even though the SME describes Lorentz violating effects, it preserves symmetry under observer rotations and boosts. Therefore, $c_{\mu \nu}$ is expected to transform like any other tensor of rank 2 (see below).
In our experiment, the two clocks have different directions of their quantization axes defined by the respective directions of the magnetic fields $\bm{^{1,2}B}$ and, correspondingly, differently oriented reference frames, i.e., clock frames CF1 and CF2.
The following procedure simplifies the analysis of the experiment:\\
(1.) We introduce an auxiliary lab frame (LF) coordinate system (origin at PTB Braunschweig, colatitude $\chi = 37.7 \degree$) with vertically upwards pointing $z$-axis, east-pointing $x$-axis, and northwards oriented $y$-axis.\\
(2.) Then, we project the quantization axes of both clocks to Earth's polar axis and determine the resulting effective colatitudes $^1\chi$ and  $^2\chi$, in our case:
\begin{equation}
^1\chi = 108.5 \degree, \, ^2\chi = 74.2 \degree.
\end{equation}

Following this procedure, we can simply adapt the same set of LF transformation formulas
to CF1 and CF2 by choosing the corresponding colatitude $^{1,2}\chi$
and an appropriate time offset, as discussed further down.

For the above introduced coordinate systems, the rotation matrix
\begin{equation}
\bm{R}=\left( \begin{array} {ccc}
-\sin(\omega T) &\cos(\omega T) & 0\\
-\cos\chi \cos(\omega T) &-\cos\chi \sin(\omega T)  &\sin\chi  \\
\sin\chi \cos(\omega T) & \sin\chi \sin(\omega T)&\cos\chi  \\
\end{array}
\right)
\label{eqR}
\end{equation}
maps $XYZ$ (SCCEF) to $xyz$ (LF) column vectors. Here, $\omega = 2\pi/(23.934~\textrm{h})$ is the sidereal day angular frequency and
$T$ modulo $2 \pi / \omega$ equals 0 whenever the $x$-axis points along the $Y$-axis (or equivalently, when the $z$-axis lies in the $X-Z$ plane).
The quantization $z$-axes of CF1 and CF2 might also deviate in east-west direction from the LF $z$-axis which we account for by determining effective longitude corrections, i.e., time offsets for the clock frames. More precisely, we
fix $T = 0$ at the moment when the CF2 $z$-axis lies in the $X-Z$ plane and points toward Sun on the 2017 vernal equinox day. This moment was in our case on March 20, 2017, at 19:00:45~UTC.
Having fixed the absolute time scale we then obtain the proper transformation equations
for CF1 by substituting $T \rightarrow T + T_{\mathrm{offset}}$ (together with $\chi \rightarrow {^1\chi}$) in all relevant expressions. For our geometry we find $T_{\mathrm{offset}} = 23.934 \, \textrm{h} \times \, 71.1\degree / 360\degree$.

Observed from the SCCEF, Earth (and with it CF1 and CF2) experiences a time-dependent boost
\begin{equation}
\bm{\beta} = \left(
\begin{array} {c}
\beta \sin(\Omega T) \\
-\beta \cos\eta \cos(\Omega T)  \\
 -\beta \sin\eta \cos(\Omega T) \\
\end{array}
\right)\label{eqb}
\end{equation}
with $\Omega = 2\pi/(365.256 \times 24~\textrm{h})$ being the sidereal year angular frequency and $\beta = 1 \times 10^{-4}$ being the magnitude of the relativistic boost.
We neglect boost contributions from Earth's daily rotation because they are two orders of magnitude smaller. This implies that the value of $T$ in $\bm{\beta}$ can be common for CF1 and CF2, i.e., ${\bm{^1\beta}} = {\bm{^2\beta}} = {\bm{\beta}}$. Note that the vernal equinox
pinning is necessary to correctly represent the phase of Earth's orbital velocity in the SCCEF.

\paragraph*{Transformation of the $c_{\mu \nu}$ tensor between clock frames and SCCEF}

Combining rotation and boost operations we can write down the explicit matrix form of the Lorentz transformation $\bm{\Lambda}$ that maps 4-vectors in the SCCEF to 4-vectors in the LF as
\begin{equation}
\bm{\Lambda} = \left(
\begin{array} {cccc}
    1        & -\beta_1  & -\beta_2 & -\beta_3\\
(-\bm{R}\bm{\beta})_1  &           &          &         \\
(-\bm{R}\bm{\beta})_2  &           &    \bm{R}     &         \\
(-\bm{R}\bm{\beta})_3  &           &          &         \\
\end{array}
\right).
\end{equation}
Due to $\beta \ll 1$ we have assumed $\gamma = 1$ for the relativistic gamma factor. Using tensor index notation, the universal transformation for any covariant tensor of rank 2, e.g. $c_{\mu \nu}$, is given by equation~\ref{LTs}. In matrix form, the symmetric and traceless $c_{\mu \nu}$ is expressed in the SCCEF as
$$   c_{MN} =
\left(
\begin{array} {cccc}
c_{TT} & c_{TX} & c_{TY} & c_{TZ} \\
c_{TX} & c_{XX} & c_{XY} & c_{XZ} \\
c_{TY} & c_{XY} & c_{YY} & c_{YZ} \\
c_{TZ} & c_{XZ} & c_{YZ} & c_{ZZ} \\
\end{array}
\right)
$$
and in the LF it reads
$$
c_{mn} =
\left(
\begin{array} {cccc}
c_{tt} & c_{tx} & c_{ty} & c_{tz} \\
c_{tx} & c_{xx} & c_{xy} & c_{xz} \\
c_{ty} & c_{xy} & c_{yy} & c_{yz} \\
c_{tz} & c_{xz} & c_{yz} & c_{zz} \\
\end{array}
\right).
$$
Rewriting equation~\ref{LTs} in matrix form leads to
\begin{equation}
c_{mn} =  (\bm{\Lambda}^{-1})^{\rm{T}} \, c_{MN} \, (\bm{\Lambda}^{-1}).
\label{transf}
\end{equation}

Adapting the LF expressions for $c_{mn}$ to CF1 and CF2 as discussed further above yields to first order in $\beta$ the following explicit time dependence for $^{2}C_{0}^{(2)}$ (see also Ref. \cite{Pruttivarasin2015}):
\begin{equation}
^{2}C_{0}^{(2)} = {^{2}A} + \sum_{j=1}^7 \left( C_j\cos(\omega_j T) + S_j\sin(\omega_j T)  \right),
\end{equation}
where $\omega_{j=1\ldots7}=\omega, 2\omega, \Omega, \Omega-\omega, \Omega+\omega, \Omega-2\omega, \Omega+2\omega$. $^{2}A$ is a constant offset and the coefficients $C_j$ and $S_j$ are listed in Table~\ref{tab2}.
For $^{1}C_{0}^{(2)}$ the expression is modified to
\begin{equation}
^{1}C_{0}^{(2)} = {^{1}A} + \sum_{j=1}^7 \left( C_j\cos(\omega_j (T+T_{\mathrm{offset}})) + S_j\sin(\omega_j (T+T_{\mathrm{offset}}))  \right).
\end{equation}
Decomposing the combined expression for ${^{1}C}_{0}^{(2)} - \, {^{2}C}_{0}^{(2)}$ into contributions from different $c_{MN}$ components and denoting the resulting time-dependent prefactors by $f_1,...,f_7$ leads to the linear fit model given in equation~\ref{fitfct}.
\begin{table*}[t]
\resizebox{\textwidth}{!}
{\begin{tabular}{cccc}
\multicolumn{1}{c}{$j$} & \multicolumn{1}{c}{$\omega_j$} & \multicolumn{1}{c}{$C_j$}& \multicolumn{1}{c}{$S_j$} \\
\hline \\
1 & $\omega$        & $-3\sin(2\chi)c_{XZ}$ &  $-3\sin(2\chi)c_{YZ}$ \\ [0.3pc]
2 & $2\omega$       & $-\frac{3}{2}(c_{XX}-c_{YY})\sin^2\chi$ & $-3c_{XY}\sin^2\chi$\\[0.3pc]
3 & $\Omega$                 &$-\frac{1}{2} \beta \left( 3\cos(2\chi) +1\right)
\left( c_{TY}\cos\eta-2c_{TZ}\sin\eta \right)$&
 $\frac{1}{2} \beta c_{TX} \left( 3\cos(2\chi) +1 \right)$\\[0.3pc]
4 & $\Omega-\omega$ & $\frac{3}{2} \beta c_{TX} \sin\eta \sin(2\chi)$ &
 $-\frac{3}{2} \beta \sin(2\chi) \left(  c_{TY} \sin\eta+c_{TZ} (1+\cos\eta)\right)$  \\[0.3pc]
5 & $\Omega+\omega$ & $\frac{3}{2} \beta c_{TX} \sin\eta \sin(2\chi)$&
$-\frac{3}{2} \beta \sin(2\chi) \left(  c_{TZ}(1-\cos\eta) - c_{TY}\sin\eta\right)$    \\[0.3pc]
6 & $\Omega-2\omega$& $-3\beta c_{TY}\cos^2(\eta/2) \sin^2\chi$&
     $-3\beta c_{TX}\cos^2(\eta/2) \sin^2\chi$  \\[0.3pc]
7 & $\Omega+2\omega$& $3\beta c_{TY}\sin^2(\eta/2) \sin^2\chi$&
 $-3\beta c_{TX}\sin^2(\eta/2) \sin^2\chi$ \\
\end{tabular}}
\caption{Amplitudes of the $^{1,2}C_{0}^{(2)}$ frequency components. For the sake of clarity, we have dropped the clock frame specifiers for the effective colatitudes $^{1,2}\chi$.}
\label{tab2}
\end{table*}

\paragraph*{Matrix element of the $T_{0}^{(2)}$ operator in Yb$^+$}

The matrix element of the $T_{0}^{(2)}$ operator in the $| Jm_{J} \rangle$ basis, $\langle Jm_{J} |\mathbf{p}^{2}-3p_{z}^{2}| Jm_{J} \rangle$ is
expressed through the reduced matrix element of the $T^{(2)}$ operator using
the Wigner-Eckart theorem (cf. Ref.~\cite{Dzuba2016})%
\begin{equation*}
\langle Jm_{J} |T^{(2)}_{0}| Jm_{J} \rangle=(-1)^{J-m_{J}}\left(
\begin{array}{ccc}
J & J & J \\
-m_{J} & 0 & m_{J}
\end{array}
\right) \langle J||T^{(2)}||J\rangle.
\end{equation*}
Explicitly we find
\begin{equation}  \label{eq10}
\langle Jm_{J} |\mathbf{p}^{2}-3p_{z}^{2}| Jm_{J} \rangle = \frac{-J\left( J+1\right)
+3m_{J}^{2}}{\sqrt{\left( 2J+3\right) \left( J+1\right) \left( 2J+1\right)
J\left( 2J-1\right)}} \, \langle J||T^{(2)}||J \rangle.
\end{equation}

The matrix elements between hyperfine states $|JIFm_{F}\rangle$,
where $\mathbf{F}$ is the total angular momentum $\mathbf{F}=\mathbf{J%
}+\mathbf{I}$, $\mathbf{I}$
is the nuclear spin, and $\mathbf{J}$ is the total angular momentum of the
electrons
are derived below. Using the Wigner-Eckart theorem and assuming that the general tensor operator $T_{q}^{(k)}$ acts only on the electronic part of the total wave function $|JIFm_{F}\rangle$, we have
\begin{equation*}
\langle J^{\prime }IF^{\prime }m_{F}^{\prime }|T_{q}^{(k)}|JIFm_{F}\rangle
=(-1)^{F^{\prime }-m_{F}^{\prime }}\left(
\begin{array}{ccc}
F^{\prime } & k & F \\
-m_{F}^{\prime } & q & m_{F}
\end{array}
\right) \langle J^{\prime }IF^{\prime }||T^{(k)}||JIF\rangle,
\end{equation*}%
where%
\begin{equation*}
\langle J^{\prime }IF^{\prime }||T^{(k)}||JIF\rangle =(-1)^{F+J'+I+k}
\sqrt{(2F'+1)(2F+1)}
\left\{
\begin{array}{ccc}
J & I & F \\
F^{\prime } & k & J^{\prime }%
\end{array}%
\right\} \langle J^{\prime }||T^{(k)}||J\rangle .
\end{equation*}%
For $T_{0}^{(2)}$ we obtain
\begin{eqnarray}
\langle J^{\prime }IF^{\prime }m_{F}^{\prime }|T_{0}^{(2)}|JIFm_{F}\rangle
&=& \delta_{m_{F}' m_{F}} (-1)^{F'- m_{F}}
\left(
\begin{array}{ccc}
F' & 2 & F \\
-m_{F} & 0 & m_{F}
\end{array}
\right) \langle J^{\prime }IF^{\prime }||T^{(2)}||JIF\rangle , \nonumber \\
\langle J^{\prime }IF^{\prime }||T^{(2)}||JIF\rangle  &=&(-1)^{F+J^{\prime}+I}
\sqrt{(2F'+1)(2F+1)}
\left\{
\begin{array}{ccc}
J & I & F \\
F^{\prime } & 2 & J^{\prime }%
\end{array}%
\right\} \langle J^{\prime }||T^{(2)}||J\rangle. \nonumber
\label{T2_hfs}
\end{eqnarray}
Converting the Yb$^+$ $4f^{13} 6s^2 \; {^2F}_{7/2}$ value $\langle J =7/2 \; ||T^2|| \; J=7/2 \rangle = -135 \; \mathrm{a.u.}$ from Ref.~\cite{Dzuba2016} to SI units and substituting it into the last equation,
we find for our $^{171}$Yb$^+$ ($I=1/2$) excited clock state
\begin{equation*}
\langle ^2F_{7/2}, F=3, m_F=0 \; | \, T_0^{(2)}/6m \, | \; ^2F_{7/2}, F=3, m_F=0 \rangle \; = \; h \times 2.6 \times 10^{16} \, \mathrm{Hz},
\end{equation*}
where $h$ is the Planck constant.
One can immediately confirm
this result by interpreting
the $| F=3, m_F=0 \rangle$ wave function as an equal superposition of the product states
\begin{equation*}
| J=7/2, m_J=-1/2 \rangle \, | I=1/2, m_I=1/2 \rangle \nonumber
\end{equation*}
and
\begin{equation*}
| J=7/2, m_J=1/2 \rangle \, | I=1/2, m_I=-1/2 \rangle. \nonumber
\end{equation*}

\paragraph*{Systematic clock frequency shifts}
The largest frequency shift results from thermal radiation emitted from the trapped ion's environment close to room temperature. The radio frequency induced temperature rise of parts of the ion trap assemblies increases the effective temperature of line-shifting thermal radiation
by 2.1 (1.3)~K and 1.5 (1.1)~K
above ambient temperature for clock 1 and 2, respectively \cite{Dolezal2015}. In addition to the effective temperature uncertainty, the uncertainty of the differential polarizability causes a $1.3\times 10^{-18}$ contribution. Since this part is common for both clocks, it does not contribute to the uncertainty of the clock difference.

Stray electric fields displace the ion from the nodal point of the trapping field causing excess micromotion (EMM) that leads to second-order Doppler and dc Stark shifts. For both systems, EMM is detected through photon correlation measurements. The magnitude is quantified by the modulation index $\beta_m$ and inferred from the relative modulation amplitude of the photon correlation signal \cite{Keller2015}. The modulation index $\beta_m$ is found to be smaller than 0.04 for clock 1 and 0.03 for clock 2 for the investigated directions of cooling laser light. As discussed in reference \cite{Dube2013},  $\beta_m$ can be used to calculate EMM induced second-order Doppler shifts of $-0.7 \times 10^{-18}$ and $-0.4 \times 10^{-18}$ and dc Stark shifts of $-0.4 \times 10^{-18}$ and $-0.3\times 10^{-18}$. Because the recorded EMM only gives an upper bound, we assign the full shift as the uncertainty. The ions' residual thermal motion after Doppler cooling and anomalous motional energy gain \cite{Ludlow2015} during the subsequent interrogation period at a rate of 40(20)~$\hbar \; {^1\omega}_r/$s and 100(20)~$\hbar \; {^2\omega}_r/$s for clock 1 and 2, however, cause the larger part of these shifts. The mean kinetic temperature of 1.0~mK and 2.2~mK during the interrogation induces second-order Doppler shifts of $-1.6 \times 10^{-18}$ and $-3.6 \times 10^{-19}$ and dc Stark shifts of $-0.8 \times 10^{-18}$ and $-1.1 \times 10^{-18}$ for clock 1 and 2, respectively. Measurements on the gain of motional energy during the interrogation were performed before and after the measurement campaign and agreed within their  uncertainties. To account for potential changes during the 6-month period, we assign 50\% uncertainties to both shifts related to the residual thermal motion of the ion.

In addition to relying on the E3 clock transition, we routinely operate both clocks for evaluation and diagnosis purposes on the electric quadrupole (E2) transition from the ground state to the $^2$D$_{3/2} \, (F=2,m_F=0)$ excited state. With
its significantly larger sensitivity to external fields, the E2 transition allows for the investigation of shift-inducing residual
electric and magnetic fields on a magnified scale. For instance, the so-called quadrupole shift that results from the interaction between residual electric field gradients and the quadrupole moment of the $^2$F$_{7/2}$ clock state is corrected for using measurements performed on the E2 transition and scaling them by the known relative sensitivity. Similarly, the second-order Zeeman shift is determined using the Zeeman splitting observed on the E2 transition \cite{Godun2014}.

The uncertainty associated with collisions with the background gas is estimated using a model based on phase changing Langevin collisions \cite{Rosenband2008}. Although the collision rate appears to be similar in both traps, we assume uncorrelated shifts in the frequency difference because of the different gas composition in the glass and the stainless steel vacuum enclosures of clock 1 and clock 2, respectively.

For stabilization of the laser light, a drift-compensating second-order integrating servo system is utilized. The corresponding servo uncertainty results from remaining non-linear frequency drifts of the reference cavity. In comparison to a previous evaluation \cite{Huntemann2016}, the longer interrogation time and the smaller nonlinear drift of the reference cavity reduce the uncertainty. Since clock 2 uses light stabilized by clock 1, potential shifts are suppressed and the servo uncertainties for clock 2 and the clock comparison are correspondingly reduced.

\textbf{Acknowledgments:} We thank Brett Altschul, Akihisa Goban, Ross Hutson, Alan Kosteleck\'{y}, Tanja Mehlst\"aubler, Matt Mewes, Arnaldo Vargas-Silva and Jiehang Zhang for helpful discussions, and Burghard Lipphardt for experimental assistance. \textbf{Funding:} This research has received funding from the European Metrology Programme for Innovation and Research (EMPIR Project OC18) co-financed by the Participating States and
from the European Union's Horizon 2020 research and
innovation program, and from DFG through CRC 1227 (DQ-mat). This work was also supported in part by the Office of Naval Research, USA, under award number N00014-17-1-2252, NSF grant PHY-1620687 (USA), and by the Russian Foundation for Basic Research under Grant No. 17-02-00216.
\textbf{Author contributions:}
Conception of the experiment and development of methods: C.S., N.H., E.P.
Design and construction of experimental apparatus: C.S., N.H., R.L, C.T. Data acquisition and analysis: C.S., N.H., R.L., M.S.S., S.G.P.
Preparation and discussion of the manuscript: C.S., N.H., R.L., C.T., E.P., M.S.S., S.G.P.
\textbf{Competing interests:} The authors declare that they have no competing interests.
\textbf{Data and materials availability:} All data underlying the study are available from the corresponding author on reasonable request. Correspondence and requests for materials should be addressed to christian.sanner@jila.colorado.edu.


\begin{thebibliography}{10}
\bibitem{FootNote1}
The displayed orbitals are pictorial representations of the angular part of
  corresponding hydrogen momentum space wave functions.
\end{thebibliography}
\end{document}